\documentclass[prd,nofootinbib,twocolumn,showpacs,preprintnumbers,floatfix,amsmath,amssymb,amsfonts]{revtex4}

\usepackage{graphicx}
\usepackage{amsmath}
\usepackage{dcolumn}
\usepackage{epsfig}
\usepackage{psfrag}
\usepackage{lineno}
\usepackage{color}

\RequirePackage{xspace}
\RequirePackage{ifthen}

\DeclareGraphicsExtensions{.epsi,.eps,.ps,.eps.gz,.ps.gz,.gif,.epsi.gz}
\newboolean{paper}
\setboolean{paper}{false}

\RequirePackage{xspace}

\usepackage{relsize}
\def\babar{\mbox{\slshape B\kern-0.1em{\smaller A}\kern-0.1em
    B\kern-0.1em{\smaller A\kern-0.2em R}}\xspace}

\def\belle{Belle\xspace}

\def\epem       {\ensuremath{e^+e^-}\xspace}

\def\kaon  {\ensuremath{K}\xspace}
\def\Kbar  {\kern 0.2em\overline{\kern -0.2em K}{}\xspace}
\def\Kb    {\ensuremath{\Kbar}\xspace}
\def\KKb   {\ensuremath{\kaon {\kern -0.16em \Kb}}\xspace}
\def\Kz    {\ensuremath{K^0}\xspace}
\def\Kzb   {\ensuremath{\Kbar^0}\xspace}
\def\KzKzb {\ensuremath{\Kz \kern -0.16em \Kzb}\xspace}
\def\Kp    {\ensuremath{K^+}\xspace}
\def\Km    {\ensuremath{K^-}\xspace}

\def\KpKm  {\ensuremath{\Kp \kern -0.16em \Km}\xspace}
\def\KS    {\ensuremath{K^0_{\scriptscriptstyle S}}\xspace} 
 
\def\KL    {\ensuremath{K^0_{\scriptscriptstyle L}}\xspace}

\def\Pbar    {\kern 0.2em\overline{\kern -0.2em P}{}\xspace}

\def\D       {\ensuremath{D}\xspace}
\def\Dbar    {\kern 0.2em\overline{\kern -0.2em D}{}\xspace}
\def\Db      {\ensuremath{\Dbar}\xspace}
\def\Dz      {\ensuremath{D^0}\xspace}
\def\Dzb     {\ensuremath{\Dbar^0}\xspace}
\def\DzDzb   {\ensuremath{\Dz {\kern -0.16em \Dzb}}\xspace}
\def\Dp      {\ensuremath{D^+}\xspace}
\def\Dm      {\ensuremath{D^-}\xspace}

\def\DpDm    {\ensuremath{\Dp {\kern -0.16em \Dm}}\xspace}
\def\Dstar   {\ensuremath{D^*}\xspace}
\def\Dstarb  {\ensuremath{\Dbar^*}\xspace}

\def\DDb     {\ensuremath{\D {\kern -0.16em \Db}}\xspace}

\def\DstarDstarb     {\ensuremath{\Dstar{\kern -0.16em \Dstarb}}\xspace}

\def\B       {\ensuremath{B}\xspace}
\def\Bbar    {\kern 0.18em\overline{\kern -0.18em B}{}\xspace}

\def\Bz      {\ensuremath{B^0}\xspace}
\def\Bzb     {\ensuremath{\Bbar^0}\xspace}
\def\BzBzb   {\ensuremath{\Bz {\kern -0.16em \Bzb}}\xspace}
\def\BzBz    {\ensuremath{\Bz {\kern -0.16em \Bz}}\xspace}
\def\BzbBzb  {\ensuremath{\Bzb {\kern -0.16em \Bzb}}\xspace}
\def\Bu      {\ensuremath{B^+}\xspace}
\def\Bub     {\ensuremath{B^-}\xspace}

\def\BpBm    {\ensuremath{\Bu {\kern -0.16em \Bub}}\xspace}
\def\Bs      {\ensuremath{B^0_s}\xspace}
\def\Bsb     {\ensuremath{\Bbar^0_s}\xspace}
\def\BsBsb   {\ensuremath{\Bs {\kern -0.08em \Bsb}}\xspace}

\def\P       {\ensuremath{P}\xspace}
\def\Pbar    {\kern 0.18em\overline{\kern -0.18em P}{}\xspace}

\def\BorBbar    {\kern 0.18em\optbar{\kern -0.18em B}{}\xspace}
\def\DorDbar    {\kern 0.18em\optbar{\kern -0.18em D}{}\xspace}
\def\KorKbar    {\kern 0.18em\optbar{\kern -0.18em K}{}\xspace}

\mathchardef\Upsilon="7107
\def\Y#1S{\ensuremath{\Upsilon{(#1S)}}\xspace}
\def\FourS{\ensuremath{\Upsilon{(4S)}}\xspace}

\def\chib#1  {\ensuremath{\chi_{b#1}\xspace}}

\newcommand{\spect}[4][]{\ensuremath{\ifthenelse{\equal{#1}{}} {} {#1\,} {}^{#2\!} {#3}_{#4}}}

\mathchardef\Deltares="7101
\mathchardef\Xi="7104
\mathchardef\Lambda="7103
\mathchardef\Sigma="7106
\mathchardef\Omega="710A

\def\Deltabar {\kern 0.25em\overline{\kern -0.25em \Deltares}{}\xspace}
\def\Lbar     {\kern 0.2em\overline{\kern -0.2em\Lambda\kern 0.05em}\kern-0.05em{}\xspace}
\def\Sigbar   {\kern 0.2em\overline{\kern -0.2em \Sigma}{}\xspace}
\def\Xibar    {\kern 0.2em\overline{\kern -0.2em \Xi}{}\xspace}
\def\Obar     {\kern 0.2em\overline{\kern -0.2em \Omega}{}\xspace}
\def\Nbar     {\kern 0.2em\overline{\kern -0.2em N}{}\xspace}
\def\Xb       {\kern 0.2em\overline{\kern -0.2em X}{}\xspace}

\newcommand{\tev}{\ensuremath{\mathrm{\,Te\kern -0.1em V}}\xspace}
\newcommand{\gev}{\ensuremath{\mathrm{\,Ge\kern -0.1em V}}\xspace}
\newcommand{\mev}{\ensuremath{\mathrm{\,Me\kern -0.1em V}}\xspace}
\newcommand{\kev}{\ensuremath{\mathrm{\,ke\kern -0.1em V}}\xspace}
\newcommand{\ev}{\ensuremath{\mathrm{\,e\kern -0.1em V}}\xspace}
\newcommand{\gevc}{\ensuremath{{\mathrm{\,Ge\kern -0.1em V\!/}c}}\xspace}
\newcommand{\mevc}{\ensuremath{{\mathrm{\,Me\kern -0.1em V\!/}c}}\xspace}
\newcommand{\gevcc}{\ensuremath{{\mathrm{\,Ge\kern -0.1em V\!/}c^2}}\xspace}
\newcommand{\mevcc}{\ensuremath{{\mathrm{\,Me\kern -0.1em V\!/}c^2}}\xspace}

\def\invfb   {\ensuremath{\mbox{\,fb}^{-1}}\xspace}

\definecolor{Red}{rgb}{1,0,0}
\definecolor{Green}{rgb}{0,1,0}
\definecolor{Blue}{rgb}{0,0,1}
\definecolor{Black}{rgb}{0,0,0}

\def\to                 {\ensuremath{\rightarrow}\xspace}

\def\pep2{PEP-II}
\def\BF{$B$ Factory\xspace}
\def\BFs{$B$ Factories\xspace}

\def\gsim{{~\raise.15em\hbox{$>$}\kern-.85em
          \lower.35em\hbox{$\sim$}~}\xspace}
\def\lsim{{~\raise.15em\hbox{$<$}\kern-.85em
          \lower.35em\hbox{$\sim$}~}\xspace}

\def\CP      {\ensuremath{C\!P}\xspace}

\def\CPT     {\ensuremath{C\!PT}\xspace}
\def\C       {\ensuremath{C}\xspace}
\def\P       {\ensuremath{P}\xspace}
\def\T       {\ensuremath{T}\xspace}
\def\TP     {{triple product}\xspace}  

 \def\p1b{\ensuremath{\phi_1}\xspace}

\xspace

\def\jetset74   {\mbox{\tt Jetset \hspace{-0.5em}7.\hspace{-0.2em}4}\xspace}

\long\def\inst#1{\par\nobreak\kern 4pt\nobreak
    {\it #1}\par\vskip 10pt plus 3pt minus 3pt}

\begin{document}


{\pagestyle{empty}

\par\vskip 3cm

\title{
\Large \boldmath
  \C, \P, and \CP asymmetry observables based on triple product asymmetries
}

\author{A. J. Bevan}
\affiliation{Queen Mary University of London, Mile End Road, E1 4NS, United Kingdom}

\date{\today}

\begin{abstract}
%
%
Triple product asymmetries have been used to probe \CP violation in $K$, \D, and \B decays.
Here we review the interpretation of those asymmetries, and note that it is possible
to construct twelve measurable triple product asymmetries for the decay of a particle into
a four body final state.
Eight of these asymmetries are introduced here and nine have never been measured before.
These can be used to systematically test \C, \P, and \CP symmetries in
decays to four body final states.  In particular we note that these asymmetries
can be used to study symmetry invariance in Higgs, $Z^0$, top-quark, and hadron decay (both
baryon and meson) as well as for $\tau^\pm$ decay.  At low energy theoretical uncertainties
arising from QCD effects will impede the interpretation of some of these asymmetries.
\end{abstract}


\maketitle

\vfill

}

\setcounter{footnote}{0}

\section{Introduction}
\label{intro}
%
%
The weak interaction of quarks and leptons is known to maximally violate parity (\P)~\cite{Wu:1957my} and 
charge conjugation (\C)\footnote{For example this is evident from meson 
decays to flavour specific final states.} symmetries.  The combination \CP is rarely
violated in weak decay, however in certain circumstances \CP violation is found
to occur~\cite{Christenson:1964fg,Aubert:2001nu,Abe:2001xe}, and that in itself has significant ramifications for trying to 
understand why the universe is matter dominated~\cite{Sakharov:1967dj}.  While \C and \P violation
is understood in terms of chiral nature of the weak interaction, \CP violation
is described by the ad-hoc $3\times 3$ unitary Cabibbo-Kobayashi-Maskawa (CKM)
quark mixing matrix~\cite{Cabibbo:1963yz,Kobayashi:1973fv}.  This matrix in turn
is related to the Higgs Yukawa couplings in the standard model of particle physics (SM) and is not yet understood at a 
fundamental level.
The use of scalar \TP asymmetries to test the weak interaction 
description of nature has been explored for many decades 
(for example see \cite{Donoghue:1987wu,Valencia:1988it,Atwood:1994kn,Datta:2003mj,Kang:2009iy,Atwood:2012ac} and references therein).
These asymmetries have been measured in kaon, charm, and 
\B decays and a detailed discussion of those results is summarised in a recent review~\cite{Gronau:2011cf}.
We revisit the interpretation of \TP asymmetries in the context
of the decay of a particle $M$ to a four body final state $a b c d = f$.  A total of twelve 
asymmetries are discussed in the context of testing weak interactions under \C, \P, and \CP
symmetry transformations, where three have been measured previously, eight 
have been introduced here, and the remaining one is mentioned in~\cite{Kang:2009iy,Gronau:2011cf}.
This paper discusses three types of decay.  The most general case (type 1)
is valid for decays where $M \neq \overline M$ and $abcd \neq \overline{abcd}$.  We consider 
two simplifications; type 2 where $M \neq \overline M$ and $abcd = \overline{abcd}$,
and type 3 where $M = \overline M$ and $abcd = \overline{abcd}$.  The twelve
asymmetries are valid for cases 1 and 2, and compactify to only a single 
unique asymmetry in case 3, which is a test of both \P and \CP.  Examples of
each of these types of decay are $D_s^+\to \KS K^+ \pi^+\pi^-$, $D^0\to K^+ K^- \pi^+\pi^-$, and
$\KL\to \pi^+\pi^- e^+e^-$, respectively.

Following this we proceed to consider the use of \TP asymmetries
to constrain symmetry non-invariance in the decay of top-quarks, $H^0$, $Z^0$, 
heavy meson systems, and for $\tau$ leptons.  These symmetry invariance tests can be performed at the
LHC general purpose detectors ATLAS and CMS, future Higgs and top factories, and flavour 
physics experiments such as \babar, Belle (II), BES III, LHCb, NA62, and VEPP.  They are
valid in a general sense, and can be used to probe the weak interaction properties
of particles in the context of the SM.  They are also applicable in the 
study of new physics, should experiments find new particles that decay into four body 
final states, at some point in the future.  Similarly one can indirectly probe for
new physics via interference effects that could result in \C, \P, or \CP violation 
in $Z^0$ or $H^0$ decay beyond the SM.

The \TP is given by $\psi = \vec{p}_c \cdot (\vec{p}_a \times \vec{p}_b)$, where 
the $\vec{p}_i$, $i=a,b,c$ are particle momentum vectors computed in the rest 
frame of $M$.  
If one considers the decay of a particle into a final state with four daughters, then one
can construct a \TP that uniquely defines the kinematics of that final state in terms of
any three of the four as the mass of the mother particle $M$ can be used to constrain the 
kinematics.  The daughter particles themselves are not required to be stable.

\begin{figure}[!ht]
\begin{center}
  \includegraphics[width=8cm]{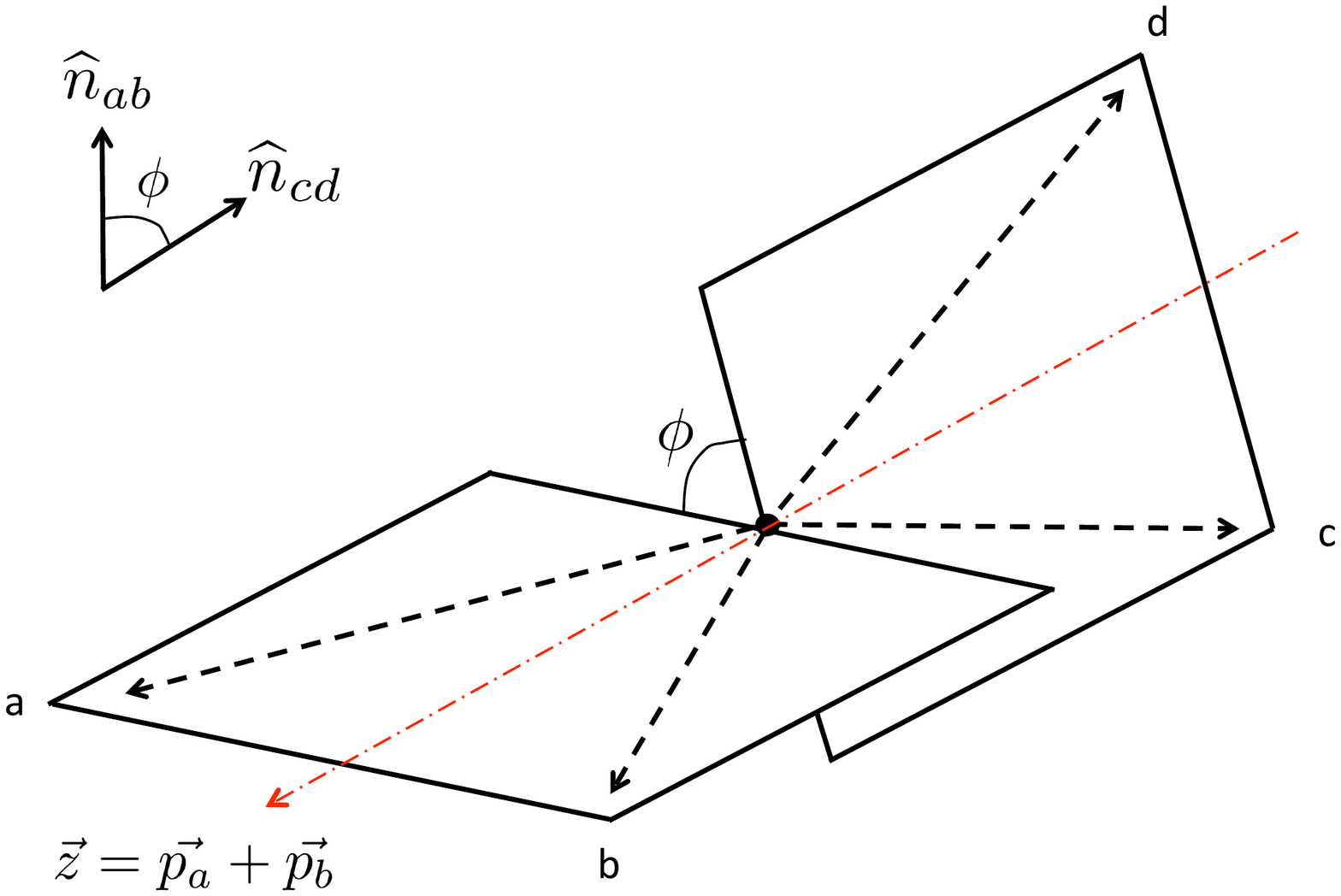}
  \caption{The reference decay $M\to f$ as described in the text.}
  \label{fig:decayschematic}
\end{center}
\end{figure}

For the four body decay $M\to f$ shown in Fig.~\ref{fig:decayschematic} one 
can define the angle between the two decay planes defined by
$ab$ and $cd$ as $\phi$ (or equivalently in terms of the normals to the decay planes).
Conventionally one defines the normals to these decay planes as $\widehat{n}_{ab}$
and $\widehat{n}_{cd}$, respectively.
It is straightforward to show that
\begin{eqnarray}
\sin \phi  &=& (\widehat{n}_{ab} \times \widehat{n}_{cd})\cdot \widehat{z}.
\end{eqnarray}
The unit vector $\widehat{z}$ is defined as the direction $\vec{p}_a+\vec{p}_b$.
Hence one can compute asymmetries based on the sign of $\psi$ of $\sin\phi$ or as a function of $\sin\phi$.
In the case of the decay $\Kz\to \gamma\gamma$ Dreitlein and Primakoff have noted that 
a \TP can be constructed to measure a time-dependent linear polarisation 
effect~\cite{Dreitlein:1961zz}.
Terms of $\sin 2\phi$ are interesting to study for some decay channels as noted
in~\cite{Valencia:1988it,Datta:2003mj,Gronau:2011cf,Bobeth:2008ij,Alok:2011gv}, where the focus is on
the study of the linear polarisation basis of $D$ and $B_{(s)}$ decays to two vector particle
states, which subsequently decay to a four particle final state.  

A \TP is even (odd) under \C (\P, \T, and \CP), as can be seen in the following:
\begin{eqnarray}
\C[\vec{p}_c \cdot (\vec{p}_a \times \vec{p}_b)] = \overline{\vec{p}_c \cdot (\vec{p}_a \times \vec{p}_b)},\\
\P[\vec{p}_c \cdot (\vec{p}_a \times \vec{p}_b)] = -\vec{p}_c \cdot (\vec{p}_a \times \vec{p}_b),\\
\T[\vec{p}_c \cdot (\vec{p}_a \times \vec{p}_b)] = -\vec{p}_c \cdot (\vec{p}_a \times \vec{p}_b),\\
\CP[\vec{p}_c \cdot (\vec{p}_a \times \vec{p}_b)] = -\overline{\vec{p}_c \cdot (\vec{p}_a \times \vec{p}_b)}.
\end{eqnarray}
It follows that one can compare event distributions for $\sin n\phi $ (or $\psi)>0$, where $n=1,\, 2$, against
those with $\sin n\phi$ or $\psi <0$ in order to probe the 
nature of the \TP under these symmetry transformations.  
However it is necessary to study physical observables directly.  One requires
the decay of some parent particle to the final state where we may subsequently construct a \TP,
and also physically determine the symmetry transformed process to compare with.
Hence it is not 
possible to test \T as one can not prepare the conjugate process $abcd\to M$.  
However, \P can be studied by comparing $M \to a b c d$ events for the triple product being 
greater than ($+$) or less than ($-$) zero and similarly for the Charge Conjugate process $\overline{M} \to \overline{f}$.  One
can also test \C and \CP using similar asymmetries as discussed below. We
denote the rates of particle (anti-particle) decay as $\Gamma_\pm$ ($\overline\Gamma_\pm$), 
where the subscript indicates the sign of the \TP.  
In the following we construct a number of asymmetry observables that can be computed as a function, 
or by integrating over positive and negative values, of the triple product.

\section{Asymmetries}
\label{asymmetries}
%
%
By considering decay rates of (anti-)particles under the Parity operator it is possible to construct 
the following asymmetries 
\begin{eqnarray}
A_{\P} = \frac{\Gamma_{+} - \Gamma_{-}} {\Gamma_{+} + \Gamma_{-}}, \,\,\,\,\,
\overline{A}_{\P} = \frac{\overline\Gamma_{+} - \overline\Gamma_{-}} {\overline\Gamma_{+} + \overline\Gamma_{-}}.\label{eq:tp:parity}
\end{eqnarray}
In the absence of final state interactions (FSI)~\footnote{For a discussion of FSI please see, 
for example, Refs~\cite{Watson:1952ji,Slobodrian:1971an}.} 
that arise from long distance strong interaction 
effects (which therefore conserve \C, \P, and \CP) a non-zero value of $A_{\P}$ or $\overline{A}_{\P}$ signifies 
Parity violation.
If FSI are present then one has to understand the impact of this on the \TP asymmetry 
in order to extract the magnitude of the underlying Parity violating effect, if that is possible.
HQET and factorisation calculations are being tested for $\Lambda_b$ 
decays with regard to the measurement of $\alpha_b$ as discussed below where it is now possible to 
compare experiment with predictions. However, in general for lower energy systems soft QCD effects are
not well understood and it is not clear when or if equivalent exercises can be made.
Discussion on the importance of these effects for different final 
states can be found in Refs~\cite{Valencia:1988it,Bauer:1986bm}, however more work in this
area might help us to interpret results that have appeared in recent years.
Another issue to be mindful of is that as strong phases can vary across phase space, if one 
integrates over part of that phase space (e.g for $\sin\phi>0$) to compare with another 
region (e.g. $\sin\phi<0$), any strong phase difference contribution to an asymmetry may 
generate an artificial signal unrelated to the weak dynamics of interest.
This can be important for $K$, $D$, and $B$ meson systems where significant FSI contributions may be
manifest in some decays.

By considering both \C and \CP operators on Eqns~(\ref{eq:tp:parity}) 
one can construct \C and \CP asymmetries from the difference and sum of 
$A_{\P}$ and $\overline{A}_{\P}$, which we denote as 
\begin{eqnarray}
a_{\C}^{\P}&=& \frac{1}{2}\left( A_{\P} - \overline{A}_{\P}\right),\label{eq:tp:cone}\\
a_{\CP}^{\P} &=& \frac{1}{2}\left( A_{\P} + \overline{A}_{\P}\right),\label{eq:tp:cpone}
\end{eqnarray}
respectively.
The literature often refers to $A_{\P} = A_T$, $\overline{A}_{\P} = \overline{A}_T$ and 
the \CP asymmetry $a_{\C}^{\P} = {\cal A}_T$ in terms of the \T-odd nature of the underlying \TP; we prefer
to denote the symmetry under scrutiny.  

%
%
Now we consider (anti-)particle decays under the operation of 
Charge Conjugation to construct the following two asymmetries
\begin{eqnarray}
A_{\C} = \frac{\overline\Gamma_{-} - \Gamma_{-}} {\overline\Gamma_{-} + \Gamma_{-}}, \,\,\,\,\,
\overline A_{\C} = \frac{\overline\Gamma_{+} - \Gamma_{+}} {\overline\Gamma_{+} + \Gamma_{+}}.\label{eq:tp:c}
\end{eqnarray}
Weak phases change sign under \C, whereas strong phases
do not, thus any strong interaction contribution 
should cancel in $A_{\C}$ and $\overline A_{\C}$.  Hence non-zero
values of these parameters signify violation of the Charge Conjugation symmetry.
On noting that the two \C asymmetries are themselves conjugated under \P, one can
combine these to form the Parity asymmetry
\begin{eqnarray}
a_{\P}^{\C} &=& \frac{1}{2}(A_{\C} - \overline{A}_{\C}), \label{eq:tp:pthree}
\end{eqnarray}
and similarly construct the equivalent \CP asymmetry
\begin{eqnarray}
a_{\CP}^{\C} &=& \frac{1}{2}(A_{\C} + \overline{A}_{\C}). \label{eq:tp:cpthree}
\end{eqnarray}
We note that on combining these terms into a single fraction the numerators for
$a_{\P}^{\C}$ and $a_{\C}^{\P}$ are equal, however the denominators differ.
The common condition for these symmetry violations is 
$\overline{\Gamma}_- \Gamma_+ - \overline{\Gamma}_+\Gamma_- \neq 0$.

%
%
Finally we can consider the decay rates under the \CP transformation, yielding
\begin{eqnarray}
A_{\CP} = \frac{\overline{\Gamma}_{+} - \Gamma_{-}} {\overline{\Gamma}_{+} + \Gamma_{-}}, \,\,\,\,\,
\overline{A}_{\CP} = \frac{\overline\Gamma_{-} - \Gamma_{+}} {\overline\Gamma_{-} + \Gamma_{+}},\label{eq:tp:cp}
\end{eqnarray}
where the sign convention is chosen to follow that used for 
direct \CP asymmetries and for time-dependent \CP asymmetries; i.e. the difference
between the normalised anti-particle and particle rates.
As with the observables $A_\P$ and $\overline A_\P$, these \CP asymmetries may not
be theoretically clean for some low energy measurements as different parts of
phase space are sampled when constructing $A_{\CP}$ and $\overline{A}_{\CP}$.
It is possible to construct two additional asymmetries by considering \P and \C 
transformations on these \CP asymmetries, which are given by
\begin{eqnarray}
a^{\CP}_{\P} = \frac{1}{2}(A_{\CP} - \overline A_{\CP}),\\
a^{\CP}_{\C} = \frac{1}{2}(A_{\CP} + \overline A_{\CP}).
\end{eqnarray}
On combining these terms into a single fraction the numerators for these observables 
are the same as those for $a_{\CP}^{\P}$ and $a_{\CP}^{\C}$, respectively, but the denominators 
are different.  The underlying conditions for this kind of symmetry violation for $a^{\CP}_{\P}$
and $a_{\CP}^{\P}$ is $\overline{\Gamma}_+ \Gamma_+ - \overline{\Gamma}_-\Gamma_- \neq 0$.
Likewise for $a^{\CP}_{\C}$ and $a_{\CP}^{\C}$ the condition is $\overline{\Gamma}_- \overline{\Gamma}_+ - \Gamma_- \Gamma_+ \neq 0$.

The twelve symmetries introduced here are valid for decays of type 1 and 2.  For type 3 transitions only the 
six asymmetries $A_{\P}$, $\overline{A}_{\P}$, $A_{\CP}$, $\overline{A}_{\CP}$, $a_{\P}^{\CP}$, and $a_{\CP}^{\P}$ remain 
non-trivial but are all equivalent.  This remaining asymmetry is a test of both \P and \CP 
in terms of the average rate $\langle\Gamma\rangle$;
%
%
\begin{eqnarray}
A_{\P, \CP} = \frac{\langle\Gamma\rangle_{+} - \langle\Gamma\rangle_{-} }{ \langle\Gamma\rangle_{+} + \langle\Gamma\rangle_{-}},\label{eq:tp:averagedratescp}
\end{eqnarray}
or equivalently using the total rates $\Gamma_\pm$.  

Thus far the focus on studies of \TP asymmetries has been in terms of searching for a violation of \CP.
In order to do this one needs to study rare processes with two or more interfering amplitudes. This
criterion is relaxed if one cares about studying \P symmetries which is maximally violated
by the nature of the weak interaction.  The study of copious Cabibbo favoured processes in order to 
understand how the weak interaction and hadronisation processes contribute to such decays is also 
of interest.  While \P should be violated for such transitions, one expects that \C and \CP violation 
would not be manifest.

From the discussion so far is not straightforward to understand how 
measurements of these asymmetries might provide useful information
about weak interactions.  
A naive example to illustrate the use of these asymmetries 
is to assume just two interfering amplitudes divided into regions of
positive and negative triple product
\begin{eqnarray}
A_+ &=& a_1 e^{i (\phi_1 + \delta_{1, +})} +  a_2 e^{i (\phi_2 + \delta_{2, +})},\\
A_- &=& a_1 e^{i (\phi_1 + \delta_{1, -})} +  a_2 e^{i (\phi_2 + \delta_{2, -})},\\
\overline A_+ &=& a_1 e^{i (-\phi_1 + \delta_{1, +})} +  a_2 e^{i (-\phi_2 + \delta_{2, +})},\\
\overline A_- &=& a_1 e^{i (-\phi_1 + \delta_{1, -})} +  a_2 e^{i (-\phi_2 + \delta_{2, -})}.
\end{eqnarray}
The amplitudes have magnitudes $a_{1,2}$ and weak and strong phases $\phi_{1,2}$ and $\delta_{1,2}$, respectively.  
The strong phases are sub-divided according to the sign of the triple product to highlight the fact
that these are generally a function of phase space.  On substituting $\Gamma = |A|^2$ into the 
asymmetry functions we find
\begin{widetext}
\begin{eqnarray}
A_P               &\propto& r\sin\Delta\phi (\sin\Delta\delta_- -\sin\Delta\delta_+) + r\cos\Delta\phi (\cos\Delta\delta_+ - \cos\Delta\delta_-)\label{eq:tripleproduct:model:ap}\\
\overline A_P     &\propto& r\sin\Delta\phi (\sin\Delta\delta_+ - \sin\Delta\delta_-) + r\cos\Delta\phi (\cos\Delta\delta_+ - \cos\Delta\delta_-)\label{eq:tripleproduct:model:apbar}\\
A^P_C             &\propto& [(2r^2\cos\Delta \phi \sin[ \Delta \delta_- - \Delta \delta_+]) + r(1+r^2) (\sin\Delta \delta_- - \sin\Delta \delta_+)]\sin\Delta \phi \label{eq:tripleproduct:model:apc}\\
A^\P_{\CP}        &\propto& (\cos\Delta\delta_- - \cos\Delta\delta_+)(r^2 (\cos\Delta\delta_- + \cos\Delta\delta_+) + r(1+r^2)\cos\Delta\phi ) \label{eq:tripleproduct:model:apcp}\\
A_C               &\propto& 2r \sin[\Delta \delta_-] \sin[\Delta \phi] \\
\overline A_C     &\propto& 2r \sin[\Delta \delta_+] \sin[\Delta \phi] \\
A^C_P             &\propto& r\left[(1+r^2) (\sin\Delta \delta_- - \sin\Delta \delta_+) + 2r \cos\Delta \phi \sin[ \Delta \delta_- - \Delta \delta_+] \right] \sin \Delta \phi\\
A^\C_{\CP}        &\propto& r\left[(1+r^2) (\sin\Delta \delta_- + \sin\Delta \delta_+) + 2r \cos\Delta \phi \sin[ \Delta \delta_- + \Delta \delta_+] \right] \sin \Delta \phi\\
A_{\CP}           &\propto& r\cos\Delta\phi (\cos\Delta\delta_+ - \cos\Delta\delta_-) + r\sin\Delta\phi (\sin\Delta\delta_+ + \sin\Delta\delta_-)\\
\overline{A}_{\CP}&\propto& r\cos\Delta\phi (\cos\Delta\delta_- - \cos\Delta\delta_+) + r\sin\Delta\phi (\sin\Delta\delta_+ + \sin\Delta\delta_-)\\
A^{\CP}_{\C}      &\propto& r\left[(1+r^2)(\sin\Delta\delta_- + \sin\Delta\delta_+) + 2r\cos\Delta\phi\sin(\Delta\delta_- + \Delta\delta_+)  \right]\sin\Delta\phi \\
A^{\CP}_{\P}      &\propto& r(\cos\Delta\delta_+ -\cos\Delta\delta_-)[r (\cos\Delta\delta_- + \cos\Delta\delta_+) + (1+r^2)\cos\Delta\phi] .
\end{eqnarray}
\end{widetext}
Here $\Delta\phi = \phi_1 - \phi_2$, $\Delta\delta_\pm = \delta_{1,\pm} - \delta_{2,\pm}$ and $r=a_1/a_2$.  
One can see from this that six asymmetries can only be non zero for $\sin\Delta\phi \neq 0$.  These are $A^P_C$, $A_C$, $\overline A_C$, $A^C_P$, $A^\C_{\CP}$, and $A^{\CP}_{\C}$.  The asymmetries $A_C$, $\overline A_C$ have the same form as time-integrated direct \CP asymmetries studied in kaon, $D$ and $B$ decays.  The remaining asymmetries can be non zero under more relaxed conditions that include non-zero strong phase differences even if weak phase differences are zero.  The expected result that ${\cal A}_T = A^P_C \propto \sin\Delta\phi$ can be seen in Eq.~(\ref{eq:tripleproduct:model:apc}).

A number of experiments have measured what we call $A_{\P}$, $\overline A_{\P}$,
and the normalised difference of the two, $a_{\C}^{\P}$.  
We believe that the interpretation of these three observables as presented
here is physically more intuitive than the traditional description, which
typically invokes \T.
This interpretation naturally leads us to introduce the additional observables outlined above.
Likewise the \T-odd \CP asymmetry of Eq.~(\ref{eq:tp:averagedratescp}) can simply be 
described as a simultaneous test of \P and \CP violation; there is no need to invoke \T (as the 
triple product is odd under both \T and \P) as pointed out some time ago~\cite{Ellis:1999xh}.

\section{Potential applications}
\label{applications}

We now turn to the question of what decays are interesting to study, but before talking about
channels in detail, we note that non-zero \CP asymmetries are the result of interference between
two or more amplitudes.  Likewise both \C and \P asymmetries should be manifest via the V-A structure 
of the weak interaction, however quark hadronisation may wash out effects of interest. 
Therefore one can study these observables in two different contexts: 
(i) to determine non-zero asymmetries where some non-trivial effect relating to the SM
is expected and subsequently try to understand that measurement, and (ii) to test the prediction of a null effect, where a departure from some
null asymmetry would indicate a previously unconsidered amplitude beating against the 
SM contribution. 
In the SM Lagrangian the quark fields transform under \C (hence \CP) and \T such that the CKM
matrix element couplings $V_{ij}$ change to their conjugates $V_{ij}^*$; i.e. weak phases
change sign under \C, \CP, and \T, whereas strong phases do not.  
Hence with an appropriate model one may determine, or constrain, ratios of 
amplitudes and phase differences from combinations of non-zero asymmetries.
The measurements made thus far in $D$ decays have been done by integrating over phase 
space without considering the underlying contributions to the four body transition.
In general a four body amplitude analysis would give more information about the 
weak interaction for these decays, however such an analysis would be challenging,
and results would be model dependent.
An alternative simplified, and model independent, approach is to measure these asymmetries as a function of 
the possible physical combinations of $m_{ab}$ and $m_{cd}$.  Both model dependent and 
model independent approaches facilitate tests of \C, \P, and \CP using triple product asymmetries.
In particular as the weak interaction maximally violates parity for quark interactions, whilst
the strong interaction conserves this symmetry, asymmetries as a function of the mass 
distributions $m_{ab}$ and $m_{cd}$ may provide an insight into the underlying dynamics of
the decay.  Indeed it is quite plausible that a detailed study may lead to one finding 
regions of phase space that show large parity violation as well as regions that conserve
parity, so that overall one finds non-maximal asymmetries when integrating over all phase space.

Having laid out the framework of asymmetries we now proceed to consider existing,
and possible future, measurements of each of the three decay types in turn.

\emph{Type 1}: Decays of the form $M \neq \overline M$ and $abcd \neq \overline{abcd}$.
The FOCUS experiment has measured \TP asymmetries for the decays 
of $D^+$ and $D^+_s$ to $\KS K^+\pi^+\pi^-$ final states~\cite{Link:2005ti}.  
The results obtained for $a_{\CP}^{\P}$ are $0.023\pm 0.062\pm 0.022$ and $-0.036\pm 0.067\pm 0.034$,
respectively.  These have been referred to as tests of 
\T violation, however we prefer to distinguish between the triple product
tests discussed here and the more formally correct treatment of \T violation tests.
See the Particle Data Group discussion of Kabir asymmetry measurements and 
the following review~\cite{Schubert:2014ska} for further discussion on \T violation.
The corresponding results from \babar for $A^{\P}_{\CP}$ 
are $(-12.0 \pm 10.0 \pm 4.6) \times 10^{-3}$ and $(-13.6 \pm 7.7 \pm 3.4) \times 10^{-3}$~\cite{Lees:2011dx}.
These experiments also measure $A_{\P}$ and $-\overline{A}_{\P}$.
FOCUS also measures the direct \CP asymmetry 
(not discussed here)~\cite{Link:2000aw} however, the 
remaining nine asymmetries discussed here were not measured by either experiment.
\babar finds non-zero values of $A_{\P}$ and $-\overline{A}_{\P}$ for 
the $D_s^+$ decay, which in general are either indicative of FSI,
a weak phase difference, or both as can be seen from Eqns~(\ref{eq:tripleproduct:model:ap}) and (\ref{eq:tripleproduct:model:apbar}).  
In the context of the SM, 
where weak phase differences in charm decays are expected to be small, 
one would conclude that this effect is the result of FSI. However 
a more general interpretation beyond the SM can not rule out large weak
phase differences in these decays.
The asymmetries measured for $D^+$ decays
are consistent with zero.  It would be interesting to see what can
be learned from the \babar, \belle, BES III, and LHCb data using the full set of asymmetry measurements
for $D_{(s)}^+$ decays, in particular for $a_\P^\C$ which is a clean test of 
Parity.

The \BFs and LHCb have measured \TP asymmetries in $B\to VV$ decays, for
example the channel $B\to \phi K^*$~\cite{Aubert:2004xc,Aaij:1668516}.  Here it is noted 
that the asymmetry $a_{\CP}^{\P}$ can be constructed from the measured 
interference terms in the angular analysis.  We note that LHCb's interpretation of their measured observables
is in terms of \T-violation, subsequently invoking \CPT.  This differs
from a more natural description in terms of a \CP symmetry test as presented here. It would be interesting to 
see a re-analysis of this type of decay in the more general context of the asymmetries discussed 
here, especially given that LHCb probably has sufficient data to extract time-dependent information
and perform a non-trivial weak phase difference measurement in this mode.  A discussion of 
the use of \TP asymmetries in $B_{(s)}\to K^*(\phi) \ell^+\ell^-$ can be found in~\cite{Bobeth:2008ij,Alok:2011gv}.

%
%
\emph{Type 2}: Decays of the form $M \neq \overline M$ and $abcd = \overline{abcd}$.
Searches for \TP asymmetries in decays of this type include 
$D^0 \to K^+K^- \pi^+\pi^-$. \babar reports results only for the
three asymmetries $A_{P}$, $-\overline{A}_{P}$, and $a_{\CP}^{\P}$,
and measures $a_{\CP}^{\P} = (1.0 \pm 5.1 \pm 4.4)\times 10^{-3}$~\cite{delAmoSanchez:2010xj}.
Likewise FOCUS measures a value of $a_{\CP}^{\P}$ consistent with zero~\cite{Link:2005ti}.
As this paper was being finalised the LHCb collaboration released 
the measurement $a_{\CP}^{\P} = (1.8 \pm 2.9 \pm 0.4)\times 10^{-3}$~\cite{Aaij:2014qwa}.
\babar and LHCb find non-zero values of $A_{\P}$ and $-\overline{A}_{\P}$ for 
this decay, which again is either indicative of FSI or a non trivial weak phase difference
indicated by Eqns~(\ref{eq:tripleproduct:model:ap}) and (\ref{eq:tripleproduct:model:apbar}). 
The SM interpretation imposes small weak phase differences in this decay,
and under this assumption the non-zero asymmetries are the result of FSI.
It would be interesting to see what can
be learned from the \babar, \belle, BES III, and LHCb data using the full set of asymmetry measurements
for $\Dz$ decays, particularly for $a_\P^\C$.  For illustrative purposes 
we proceed to interpret the 
published data from \babar in terms
of the asymmetries discussed here.  The results of this are summarised in
Table~\ref{tbl:babarneutralmode}, where the published asymmetries are taken verbatim 
from~\cite{delAmoSanchez:2010xj} and error propagation, under the assumption of Gaussian 
errors, is used to compute the statistical uncertainties on the other nine quantities.  
These data are compatible with \P and \C violation, although this is not maximal.  The observed 
levels of \P and \C violation balance such that \CP is found to be conserved.  
This interpretation neglects systematic uncertainties, however we don't expect 
a significant change to the conclusions drawn if \babar re-analyse their data
in this context.  The LHCb data, when re-analysed in the same way, are completely consistent
with this interpretation of the \babar result.  The LHCb experiment
has also measured triple product asymmetries for $B_s\to \phi\phi$ decays, which are
compatible with the hypothesis of \CP conservation~\cite{Aaij:2012ud}.  A more
comprehensive study of that final state is called for. 

\begin{table}[!ht] \begin{center}
\caption {Computed asymmetries for $D^0 \to K^+K^- \pi^+\pi^-$ from the \babar data. Only 
statistical uncertainties are shown and the numbers in parentheses correspond to the
estimated statistical significance of the specified asymmetry to be non-zero in terms
of Gaussian standard deviations.  The numbers in square brackets correspond to the significance
including systematic uncertainties reported recently by \babar in~\cite{ckm2014}.\label{tbl:babarneutralmode}}
\renewcommand{\arraystretch}{1.2}
\begin{tabular}{c|l} \hline\hline
Asymmetry              & $D^0 \to K^+K^- \pi^+\pi^-$ \\ \hline
$A_{\P}$               & $-0.069 \pm 0.007  \, (9.8)\,$ $[7.5]$\\
$\overline{A}_{\P}$    & $\phantom{-}0.071 \pm 0.007 \,(10.1)\,$ $[8.8]$\\
$a_{\C}^{\P}$          & $\phantom{-}0.001 \pm 0.005 \,(0.2)\,$ $[0.2]$\\
$a_{\CP}^{\P}$         & $-0.070 \pm 0.005   \, (14.0)$ $[13.5]$\\
$A_{\C}$               & $\phantom{-}0.060 \pm 0.007 \,(8.6)\,$ $[8.3]$\\
$\overline A_{\C}$     & $-0.079 \pm 0.007 \, (11.3)$ $[10.8]$\\
$a_{\P}^{\C}$          & $\phantom{-}0.070 \pm 0.005 \, (14.0)$ $[13.5]$\\
$a_{\CP}^{\C}$         & $-0.009 \pm 0.005  \,(1.8)\,$ $[1.8]$\\
$A_{\CP}$              & $-0.008 \pm 0.007  \,(1.1)\,$ $[1.0]$\\
$\overline{A}_{\CP}$   & $-0.010 \pm 0.008  \,(1.3)\,$ $[1.1]$\\
$a^{\CP}_{\P}$         & $\phantom{-}0.001\pm 0.005 \, (0.2)\,$ $[0.2]$\\
$a^{\CP}_{\C}$         & $-0.009\pm 0.005  \, (1.8)\,$ $[1.8]$\\
\hline\hline
\end{tabular}\end{center}\end{table}

%
%
\emph{Type 3}: Decays of the form $M = \overline{M}$ and $abcd = \overline{abcd}$. These include
measurements of asymmetries in $K_{S,L}\to \pi^+\pi^-e^+e^-$~\cite{Abouzaid:2005te,Lai:2003ad}.  Here we 
interpret the non-zero asymmetry measurements as observations of both \P and \CP violation 
in the corresponding decays, which is a slightly different interpretation than the 
usual nomenclature in the literature, where they are often referred to as being
simply \T-odd \CP violating asymmetries.   KTeV and NA48 report non-zero values
of the \P and \CP violating \TP asymmetry of Eq.~(\ref{eq:tp:averagedratescp}) for the \KL
mode and NA48 find no significant asymmetry for the \KS mode.
The decays $K_{S,L}\to 4\ell$ proceed via an intermediate $\gamma^*\gamma^*$
state~\cite{Miyazaki:1974qi,Uy:1990hu,Uy:1990is,Gronau:2011cf} where one expects contributions
from both \CP conserving and \CP violating form factors.  Hence measurements
of the asymmetry given by Eq.~(\ref{eq:tp:averagedratescp}) for these modes are also of interest.

The same logic used here in the context of meson decay can be 
applied to baryon decays to four particle final states, for example
see~\cite{Bensalem:2002pz,Bensalem:2002ys,Kang:2010td}, which follows on from early studies of 
Parity violation in $\Lambda^0 \to p\pi^-$~\cite{Cronin:1963zb}.
BES III can explore these asymmetries in $\Lambda_c$ decay.
More recently the LHC has measured the corresponding parameter
$\alpha_b$ for $\Lambda_b^0 \to J/\psi \Lambda^0$ decays~\cite{Aaij:2013oxa,Aad:2014iba},
obtaining results compatible with zero from samples with event yields of 
about 1400 to 7200 events.  One could measure the twelve asymmetries 
discussed here in this decay.  Using the published ATLAS data as an example,
one could obtain statistical uncertainties of ${\cal O}(3.8)\%$ for the asymmetries 
outlined here.  Such measurements could be used to test QCD calculations to complement
existing work, and may help us approach the problem of understanding measurements made
with lower energy systems. Related to this it is worth noting that 
Gardner and He~\cite{Gardner:2013aiw} discuss the use of
triple products as a tool to study radiative $\beta$ decay.

When one considers applications in the high energy limit it is clear that one 
can test the combined \P and \CP asymmetry of Eq.~(\ref{eq:tp:averagedratescp}) 
in the decays of $Z^0$ and $H^0$ bosons 
to four particle final states (e.g. $4j$, $2j2\ell$, $4\ell$).  ATLAS have
recently reported a measurement of the branching fractions for $Z^0\to 4\ell$~\cite{Aad:2014wra}
using a few hundred events, which should be sufficient to start performing the
asymmetry measurements proposed here. 
Asymmetries for $Z\to 4j$ final states ($j$=jet)  are expected to be 
zero in the SM (see \cite{Nachtmann:1998va,Nachtmann:1999ys} and references therein).
However, this final state is more challenging than the $4\ell$ one.
Similarly one expects zero asymmetries for $H^0$ decay in the SM
and so one can use triple products to search for NP~\cite{Delaunay:2013npa,Bhattacharya:2014rra,Curtin:2013fra}.
Table~\ref{tbl:boson} summarises the expected precision on the asymmetry given by Eq.~(\ref{eq:tp:averagedratescp}) using
the LHC for $Z$ and $H$ decays to $\mu\mu ee$ final states.  We expect that the high luminosity upgrade of
the LHC (HL-LHC) could yield
statistical uncertainties of 1-2\% for these channels.
A recent paper discusses the use of triple product asymmetries to test for non-SM physics via \CP violation 
in $W\!H$ decays~\cite{Delaunay:2013npa}.  In general the associated production of a Higgs 
boson via $V\!H$ with the vector and Higgs decaying into two particle states falls into either 
a type 2 or a type 3 decay.  The process $W^\pm H^0 \to \ell^\pm \nu b\overline{b}$ 
is of the former type, while $Z\!H \to (\ell^+\ell^-, \nu\overline{\nu}) b\overline{b}$ decays 
falls into the latter category.  
It is expected that the ILC will be able to produce large samples of $ZH$ decays at both 
250 GeV and 500 GeV~\cite{tanabe:ipa}.  Based on these estimated yields, assuming a modest
reconstruction efficiency for $Z\to \ell^+\ell^-$ and $H\to b\overline{b}$, we estimate that
it will be possible to obtain precisions of 2.1 (3.7)\% and 2.5 (4.7)\% on the asymmetry of 
Eq.~(\ref{eq:tp:averagedratescp}) using inclusive (exclusive) $Z$ decays to lepton pairs
at 250 and 500 GeV, respectively.  The proposed Chinese Higgs Factory (CEPC) is expected to accumulate
about 1 million ZH decays~\cite{cepc}.  The proposed Future Circular Collider \epem variant, the so called FCC-ee 
project would be able to perform these measurements as well. 
Samples of a million $ZH$ decays could produce results about five times more precise than the ILC.
These are interesting channels to study during run 2 and during the HL-LHC assuming that systematic effects
can be controlled at the LHC. The recent results from ATLAS and CMS for this process can be found in Refs~\cite{Chatrchyan:2013zna,Aad:2014xzb}.
The ILC can also be used to make some of these measurements, where a key difference is that
the \epem production mechanism at the ILC leads to a different set of systematic uncertainties
that affect the measurements.  Hence the two types of facility will complement each other.

\begin{table}[!ht] \begin{center}
\caption {Estimated precisions on the asymmetry of Eq.~(\ref{eq:tp:averagedratescp}) for 
$Z^0$ and $H$ decays to $\mu^+\mu^- e^+e^-$ for ATLAS and CMS at the LHC and HL-LHC. 
These estimates are based on the published run 1 yields,
assuming that the asymmetry is zero.\label{tbl:boson}}
\renewcommand{\arraystretch}{1.2}
\begin{tabular}{c|cc} \hline\hline
Data sample                  & $Z^0\to \mu^+\mu^- e^+e^-$      & $H\to \mu^+\mu^- e^+e^-$ \\ \hline
Run 1 $(\sim 25 \invfb)$     & 0.12       & 0.38 \\
Run 2 $(\sim 125 \invfb)$    & 0.04       & 0.11 \\
Run 3 $(\sim 300 \invfb)$    & 0.03       & 0.07 \\
HL-LHC $(\sim 3000 \invfb)$  & 0.01       & 0.02 \\
\hline\hline
\end{tabular}\end{center}\end{table}

In the SM one expects \CP violation in the top quark sector to be small, 
because the $t\to Wb$ amplitude dominates in the decay.  Hence any large \CP violation observed
would be from physics beyond the SM (see for example~\cite{Atwood:2000tu}).
If one considers top quark decay to $Zq$, $\gamma q$ or $Wb$, it is clear that 
four particle final states are not copious, but that one can consider possible
future measurements using rare decays.  These include $\ell^+\ell^- q$
final states from $(Z,\gamma) q$ loop processes, where the $q$ hadronises and subsequently 
decays into a two body final state. The $Wb$ decays typically 
produce 3 or 5 prong final states, hence this methodology could only be 
applied to the 5 prong scenario by reconstructing, for example, a secondary $W$
boson from its decay products.  The possibility of 
performing such measurements depends on the integrated data
samples, and ability to distinguish signal from background at the LHC 
and future $e^+e^-$ colliders running at or above top production threshold.  
A lepton collider has a distinct advantage over the LHC when studying the weak 
structure of top quark decays for final states including neutrinos as one has
an additional experimental constraint: the centre of mass energy for the 
interaction is known. Top-quark decays are of type 1 or 2,
hence there are twelve observables to measure in order to probe the structure of
these weak decays.  
The down side to the observables introduced here
is the familiar penalty of requiring the study 
of rare processes. The potential benefit of doing this with top quarks is that that the rare processes
may be sensitive interferometers for new physics at higher scales.
Kiers et al. have discussed the use of triple product asymmetries to search for NP
in the three body decay $t \to b \overline{b} c$~\cite{Kiers:2011sv}.
The asymmetries introduced here complement the proposal 
of Ref.~\cite{Gedalia:2012sx} to study \CP violation in $b$ quark interactions from samples 
of top events.  The main difference here is that 
we are not restricted to studies of how a known low energy effect manifests 
at a higher scale via a tree level cascade of decays from $t$ to a final state with di-leptons. 
These asymmetries can also be applied to loop processes, once they become experimentally
accessible. It is expected that the Jarlskog invariant does not change
significantly as one runs from the $b$ mass to the Plank scale~\cite{Perez:2014nqa},
a prediction which requires precision measurements from HL-LHC
and future high energy \epem colliders to test.  A corollary of the dominant $t\to Wb$
transition is that one could test \C, \P, and \CP using top quark decay with these 
asymmetries as a test of the weak interaction at this energy scale.

The measurement principle outlined here requires that one can determine the centre of 
mass frame of the decaying particle $M$ to avoid diluting asymmetries.  
In the case of $\tau$ leptons this is
complicated by missing energy in the final state, however this issue can be
overcome in $e^+e^-\to \tau^+\tau^-$ decays at threshold, where 
the collision energy is well known from the machine.  Hence it is possible to
study these \TP asymmetries for $\tau$ leptons decaying to four particles
in the final state, and in particular probe \C, \P, and \CP violation via their weak 
decay.  One should take care with final states involving hadronic contributions in order 
to disentangle the weak effects of interest from any long distance hadronic 
contributions.  As many $\tau$ decays have kaons in the final state, which themselves
manifest \CP violation, one should take care to study modes that are theoretically and experimentally
well controlled from the perspective of asymmetry measurements. 
The twelve asymmetry observables presented here complement existing techniques
that have been used by CLEO and the \BFs to search for \CP violation in $\tau$ decay~\cite{Bonvicini:2002,Inami:2002ah,Bischofberger:2011pw,BABAR:2011aa}
and discussed in~\cite{Kiers:2008mv}.
It may be possible to perform similar measurements above production threshold, however
the efficacity of such a study would depend on the extent by which the asymmetry would
be diluted.

%
%
Up until now the decay rates discussed have been in terms of time-integrated 
quantities.  However, one can generalise the discussion to time-dependent
asymmetries given a suitable model for the decay.  For example if one considers
$\Bz_d$ decays to final states containing two pseudoscalars or one pseudoscalar 
and a vector particle the time-dependence is given by
\begin{eqnarray}
\Gamma_\pm &\propto& \eta_{\CP} (1 + C_\pm \cos\mathrm\Delta m \mathrm \Delta t - S_\pm \sin\mathrm\Delta m \mathrm \Delta t),\label{eq:tp:timedepgamma}\\
\overline \Gamma_\pm &\propto& \eta_{\CP} (1 - C_\pm \cos\mathrm\Delta m \mathrm \Delta t + S_\pm \sin\mathrm\Delta m \mathrm \Delta t).\label{eq:tp:timedepgammabar}
\end{eqnarray}
Here $\eta_{\CP} = \pm 1$ is the \CP eigenvalue of the decay of interest, $\Delta m$
is the neutral meson mixing frequency, $\mathrm \Delta t$ is the proper time measured
for the decay to a \CP eigenstate, relative to the flavour tag filter event. \footnote{This time-dependence
is for correlated production of \B meson pairs at a \BF, and the corresponding 
form in terms of the proper time $t$ is used for uncorrelated production for measurements 
at LEP and the LHC.} The coefficients
$S_\pm$ and $C_\pm$ are related to $\lambda = (q/p)(\overline A/A)$, where $q$ and $p$ parameterise
mixing in the neutral meson system and $\overline A/A$ is the ratio of decay amplitudes
for the anti-particle to particle decay transition.  The subscript sign convention
follows that of the rates used for the time-integrated quantities.
If one substitutes Eqns~(\ref{eq:tp:timedepgamma}) and (\ref{eq:tp:timedepgammabar}) into the 
previously defined asymmetries then one finds that they are all non-trivial in general.
Thus, in principle, one may be able to relate any non-zero asymmetries to $\lambda$ and 
in turn probe the underlying mixing $(q/p)$ and decay amplitude $(\overline A/A)$ structure.
The latter being defined in terms of the CKM phase structure.
It is straightforward to extended this for
$D$ and $B_s$ mesons where the lifetime difference $\mathrm \Delta \Gamma \neq 0$.  The corresponding
treatment for decays to final states with two vector particles is a little more complicated, but 
follows the same logic.
As an example one can probe \C and \P symmetry violation through measurements of the unitarity
triangle angle $\beta$ using $b\to c\overline{c}s$ transitions to 
dis-entangle the individual contributions to the overall \CP asymmetry observed in decays
such as $\Bz \to J/\psi K^*$ decays at the \BFs and the LHC. 
While \CP violation is expected to be small in
$\B_s \to J/\psi \phi$, and it will be some time before experiments approach the sensitivity
required to observe SM levels of \CP violation, it is possible to also explore
\C and \P violation in these decays at the LHC.  One of these new observables may 
permit a non-zero value of $\beta_s$ to be made before the traditional \CP measurement does,
however that possibility needs to be explored in more detail.
It should be
noted that if one combines this approach with the methodology in~\cite{Banuls:2000ki,Bevan:2013rpr}, it is possible
to perform a self-consistent set of \C, \P, \T, \CP, and \CPT tests at \babar, Belle, and
Belle II. Such measurements could be used to over-constrain the set of weak interaction symmetry 
violation combinations to fully elucidate the corresponding nature of neutral \B decays using
the entangled states produced via \FourS decay.  Likewise for charm at the $\psi(3770)$.

%
%
One can also consider the use of \TP asymmetries as a tool to study \C, \P,
and \CP symmetry violation in the decay of a number of possible new particle
states that may be found in experiments searching for physics beyond the SM.
Again the general requirement is that one has a four body decay, where one knows
the mass of the decaying particle and computes three of the four final state
particles in the rest frame of the decaying parent.  Some examples of what one may 
wish to study are portal models (i.e. dark forces~\cite{ArkaniHamed:2008qn,Batell:2009yf,Batell:2009di,Bjorken:2009mm,Essig:2009nc}), light ($<10\gev$) Higgs particles or light dark matter decays~\cite{McElrath:2005bp,Gunion:2005rw,Dermisek:2006py},
and SUSY or Exotics searches at higher energies where one may obtain a cascade of 
particles decaying to final states with jets and/or leptons~\cite{Kittel:2009fg}.  For dark sector 
searches the decay of a dark photon $A^\prime$ to a $4\ell$ ($\ell=$ charged lepton)
final state (which would be rare relative to a $2\ell$ state),
or the Higgsstrahlung process $h^\prime \to 2A^\prime \to 4\ell$ are candidates 
of the third class of decay. Regarding searches for 
physics beyond the SM being performed by the ATLAS and CMS experiments 
at the LHC, it is worth noting that many of these involve decays to four body 
final states.   Another example is the use of triple product asymmetries to search for
\CP violation in $e^{+} e^{-} \to t \bar{t} H^0$, which can be large in the two Higgs double model~\cite{BarShalom:1995jb}.

%
%
Experiments should consider revisiting analyses of \TP asymmetries in 
order to systematically understand symmetry violation in weak decay in terms of 
the Charge Conjugation, Parity, and \CP operators.  It would be interesting to see
if the use of the new observables introduced here provide additional insights 
on the nature of weak interactions and how to relate the behaviour observed in
data from the underlying weak interaction to the fully hadronised final state.

\section{Summary}
\label{summary}

%
%
In summary the physical interpretation of \TP asymmetries has been reviewed, where
we note that these are restricted to tests of \C, \P, and \CP.  There are twelve
asymmetries for decay types 1 and 2.  Only three of these twelve asymmetries have
been measured before.  Six of the asymmetries can only be non zero for a non-vanishing weak 
phase difference between interfering amplitudes, which is five more than previously noted.
Therefore it would be interesting to see what can
be learned from the full set of observables in previously measured charm and 
\B decays at the flavour experiments as well as for the other systems discussed here.
Table~\ref{tbl:babarneutralmode} is the result of reanalysing the data from \babar
for $D^0 \to K^+K^- \pi^+\pi^-$, neglecting systematic effects.  This is an
illustration of the application of the full set of \TP asymmetries discussed 
herein, where \C and \P violation is manifest.
For decays of type 3 these twelve asymmetries compactify into
a single quantity that is a simultaneous test of both \P and \CP.
 We have discussed how these asymmetries can be studied using decays of 
$H^0$ and $Z^0$ bosons as well as for quark flavour transitions, either via direct
decay in the case of the top quark, or via hadrons for the lighter quarks.
Triple product asymmetries can be used to study \C and \P violation and
to search for \CP violation in $\tau^\pm$ decay in $e^+e^-$ colliders with a centre
of mass energy corresponding to $\tau^+\tau^-$ threshold.
In the event that new particles (for example SUSY or dark-sector) were to be found, 
then the symmetry violation structure of the decay those states could be probed using 
the asymmetries discussed here.  

Given that it has been 50 years since the discovery of \CP violation and that 
we are still unable to account for the universal matter-antimatter asymmetry within
the SM, it would seem prudent to perform systematic searches for other manifestations
wherever possible, as well as ancillary measurements that may elucidate our understanding
of the origin of this phenomenon.  The \TP asymmetries discussed in this paper can be used to 
search for and study \C, \P, and \CP violation in the decay of bosons, quarks, 
and charged leptons.

This work was supported in part by National Science Foundation Grant No. PHYS-1066293 and 
the hospitality of the Aspen Center for Physics.  The author would also like to thank 
Mike Roney and Mike Sokoloff for useful discussions during the preparation of this paper.

\end{document}